\documentclass[aps,prc,twocolumn,groupedaddress,floats,epsfig,amsmath]{revtex4}

\newcommand{\la}{\left \langle}
\newcommand{\ra}{\right \rangle}
\newcommand{\pt}{p_\perp}
\newcommand{\R}{{\cal R}}
\renewcommand{\P}{{\cal P}}

\usepackage{epsfig}
\begin{document}

\title{The freeze-out mechanism and phase-space density  in \\
ultrarelativistic heavy-ion collisions}

\author{Boris Tom\'a\v sik and Urs Achim Wiedemann}
\affiliation{CERN, Theory Division, CH-1211 Geneva 23, Switzerland}

\date{June 16, 2003}

\begin{abstract}
We explore the consequences of a freeze-out criterion for
heavy-ion collisions, based on pion escape
probabilities from the hot and dense but rapidly expanding
collision region. 
The influence of the expansion and the scattering rate 
on the escape probability is studied. The 
temperature dependence of this scattering rate favors a low
freeze-out temperature of $\sim $100~MeV. In general, our
results support freeze-out along finite four-volumes rather
than sharp three-dimensional hypersurfaces, with high-$p_\perp$
particles decoupling earlier from smaller volumes. We compare our 
approach to the proposed universal freeze-out criteria using the 
pion phase-space density and its mean free path.
\end{abstract}

\maketitle
 \vskip 0.3cm

\section{Introduction}
\label{sec1}

Experiments with nuclear collisions at ultrarelativistic energies
are performed with the aim of studying strongly interacting matter
under high temperature and density. Such conditions are created 
very early in the collision. 
Hadrons at low transverse momentum form the bulk of the particles
produced in these collisions. They are
produced at the end of the hot and dense partonic phase, 
but they subsequently scatter in
the confined hadronic phase prior to decoupling from the collision
system (``freeze-out''). In general, the collective evolution of the
hot and dense matter leaves a distinct imprint on the 
phase space distribution of the fireball at freeze-out. For
example, pressure gradients generated in the early stage of the
collision and indicative of the equation of state result in 
collective transverse, radial and elliptic flow which shape
important features of hadronic one- and two-particle spectra.
However, to disentangle such information from features generated
during freeze-out a refined understanding of the decoupling process is needed. 
This is the condition which has to be satisfied locally for
decoupling to take place.

Freeze-out is often modeled as a sudden breakup of the fireball on a 
fixed three-dimensional hypersurface \cite{Cooper:1974mv}, specified
e.g. by a critical freeze-out temperature or density. This implements
in a very simplified way the physical picture that the hadronic
scattering rate drops with particle phase space density and determines 
freeze-out. However, the current difficulties in interpreting the HBT 
spectra measured at RHIC \cite{prattqm2002} motivate to go beyond a sharp
three-dimensional implementation of freeze-out.
Moreover, the recently observed 
increase in average pion phase space density from SPS 
($\sqrt{s} = 17\, A\mbox{GeV}$) to RHIC
($\sqrt{s} = 130\, A\mbox{GeV}$)~\cite{lray,cramer} indicates that---in 
contrast to an earlier suggestion~\cite{Ferenc:1999ku}---additional 
factors beyond the particle phase space density must be taken into 
account. This prompts us in the present work to discuss quantitatively
how hadrochemical composition~\cite{nagamiya,ceres}, local
temperature and collective velocity gradients~\cite{garpman}, 
and particle momentum influence the decoupling of particles
from the collision region. Since hadrochemistry does, and freeze-out
temperature or velocity gradients may change in going from SPS
to RHIC energies, this study also addresses the possible dependence 
of hadronic freeze-out on bombarding energy.

Clearly, in the general case, particles decouple from 
the fireball gradually \cite{Grassi:1994ng,Sinyukov:2002if}, as it 
is naturally implemented e.g. in cascade generator simulations. The
question arises to what extent this generic behavior can be described
by the limiting case of a sharp three-dimensional freeze-out. To
discuss this question, we first introduce in Section \ref{sec2}
the so-called escape probability which characterizes the 
probability of a particle to decouple from the system.
In Section \ref{sec3} we demonstrate on simple examples how the 
escape probability is related to the expansion strength and the 
scattering rate. This illustrates the old suggestion \cite{garpman},
that freeze-out happens when the dilution rate becomes comparable or 
larger than the scattering rate. Moreover, the scattering rate 
depends on densities of individual species and the corresponding 
cross-sections \cite{nagamiya,ceres}. For pions, e.g., scattering
on nucleons is much more important than scattering on other 
pions due to the larger cross-section of the former. In Section
\ref{sec4}, we discuss the impact of the hadrochemical composition
on the scattering rates and freeze-out at SPS and RHIC. This also
allows to address the observed change in freeze-out phase space 
density~\cite{lray,cramer}. Finally, we turn to the dependence
of the scattering rate and escape probability on temperature. 
In doing so, we find that $\approx 50\%$ of the particles are 
emitted at temperatures below $\sim 100$ MeV.
We discuss this result, together with other 
findings  of our calculations in Section \ref{sec5}.

\section{The escape probability}
\label{sec2}
The following discussion of the freeze-out process 
\cite{Grassi:1994ng,Sinyukov:2002if} is based on the
particle escape probability ${\cal P}(x,p,\tau)$ defined as
\begin{equation}
f_{\rm esc}(x,p,\tau) = {\cal P}(x,p,\tau) f(x,p,\tau)\, .
\label{e:Pdef}
\end{equation}
Here, $f(x,p,\tau)$ is the distribution function of a given particle
species and $f_{\rm esc}(x,p,\tau)$ denotes the distribution of that
fraction of particles which have decoupled from the system prior to time
$\tau$ and will not rescatter in the future. For a particle with
momentum $p$ at space-time point $(x,\tau)$, the probability to
escape the medium without future interaction is 
\cite{Grassi:1994ng,Sinyukov:2002if}
\begin{equation}
\label{e:escprob}
     {\cal P}(x,p,\tau) =
     \exp \left ( - \int_\tau^\infty \, d\bar{\tau} \,
         \R(x+v\bar{\tau},p) \right )\, .
\end{equation}
Here, the scattering rate $\R(x,p)$ denotes the inverse of the mean time
between collisions for a particle at position $x$ with momentum $p$.
The {\em opacity integral} in \eqref{e:escprob} determines the average number
of collisions of the particle after time $\tau$. Thus, 
for a particle which tries to escape the medium at time $\tau$
with velocity $v$, eq. (\ref{e:escprob}) determines the probability
that the particle succeeds to do so.

The limiting case of a sharp freeze-out along the three-dimensional 
proper time hypersurface given by $\tau_{\rm fr}$ \cite{Cooper:1974mv}
can be characterized by 
an abrupt change of the particle escape probability
(\ref{e:escprob})  from zero to one  
\begin{eqnarray}
   {\cal P}(x,p,\tau)^{\rm 3-dim.}_{\rm freeze-out}\, =  
          \left\{ \begin{array} 
                  {r@{\qquad  \hbox{for}\quad}l}                 
                  0 
                  & \tau < \tau_{\rm fr}\, , \\ 
                  1
                  & \tau > \tau_{\rm fr} \, .
                  \end{array} \right.
  \label{sharp}
\end{eqnarray}
Freeze-out along more general 
three-dimensional hypersurfaces \cite{Cooper:1974mv} can be defined 
by requiring this kind of threshold behavior of the escape probability
along those hypersurfaces. It corresponds to a scattering rate 
which changes abruptly from a large value to zero along the
freeze-out hypersurface.

Hypersurfaces of the type \eqref{sharp} are characterized by a criterion
which tests {\em only the medium}, but does not depend on the particle momentum
and particle properties. 
In the general case of continuous freeze-out, the situation is different since
freeze-out does depend on particle momentum and properties. 
To characterize the four-volume from which particles decouple, one
requires \cite{Sinyukov:2002if}
\begin{equation}
  p^\mu \partial_\mu \P(x,p,\tau) > 0 \, .
  \label{e:ufc}
\end{equation}
This specifies the region in which the amount of escaped particles of 
momentum $p^\mu$ is growing. Note that this general condition also 
characterizes the freeze-out hypersurface in case of the sharp 
freeze-out according to eq.~\eqref{sharp}. Condition (\ref{e:ufc})
assumes, however, that there is no additional particle production, i.e.,
$p^\mu \partial_\mu f(x,p,\tau) = 0$. In the presence of such particle
production, one generalizes \eqref{e:ufc} to $\partial_\mu (p^\mu \P f) > 0$.


In general, the evaluation of the criterion \eqref{e:ufc} is
complicated. In the Discussion, we shall consider a related,
much simpler condition which gives some access to the structure of
a four-dimensional freeze-out region on the basis of the escape 
probability.

\section{The dependence of particle freeze-out on expansion}
\label{sec3}

It has been argued a long time ago \cite{garpman} that 
freeze-out happens when the dilution rate becomes 
larger than the collision rate. Here, we demonstrate how the collision rate
and the expansion strength determine the escape probability.

To this end, we consider the simplified case of a particle
of vanishing transverse momentum in the center of the fireball.
Such a particle  does not propagate
through layers of different density but---due to expansion---finds 
itself in a medium of decreasing density. This simplification
allows to illustrate the effect of dynamical properties of the 
collision region on freeze-out without being sensitive to further
complications of the general case such as finite-size effects
which depend on the production point and velocity of the 
test particle. We start the discussion of the escape probability
\eqref{e:escprob} by considering a simple power-law ansatz for 
the scattering  rate
\begin{equation}
\label{rtau}
\R(p=0,\, r=0,\,\tau) = 
\R(\tau) = \R_0 \left ( \frac{\tau_0}{\tau} \right )^\alpha \, ,
\quad \alpha > 1\, ,
\end{equation}
Here, $\R_0$ is the scattering rate at the time $\tau_0$. Below,
we detail soon the assumptions on which (\ref{rtau}) is based and 
how the exponent $\alpha$ characterizes the expansion strength
of the system. To set the stage, however, we consider first
the opacity integral of (\ref{rtau}),
\begin{equation}
\label{e:opac}
\int_{\tau_0}^\infty d\tau\, \R(\tau) = \frac{\R_0 \tau_0}{\alpha - 1} \, .
\end{equation}
This illustrates the typical interplay of scattering and expansion: 
a given escape probability can be obtained for different values of
the scattering rate, with a higher scattering rate $\R_0$ compensated 
by a stronger expansion (larger $\alpha$). 

We further illustrate this point with two analytically accessible
models for the fireball expansion. Both are based on a factorized ansatz
for the scattering rate in terms of the (averaged) cross-section 
$\sigma$ for scattering in the medium, and the average velocity 
$\bar v_{\rm rel}$ relative to other particles,
\begin{equation}
\label{e:rate}
     \R(\tau) = \sigma \rho(\tau) \bar v_{\rm rel}\, .
\end{equation}
While the time dependence of $\R$ does not factorize in the
general case, one may hope to capture the dominant features of
a realistic dynamical evolution by retaining the time dependence
of the density, only. 

For the first model, we choose a power-law fall-off of the density
\begin{equation}
\rho(\tau) = \rho_0 \left ( \frac{\tau_0}{\tau} \right )^\alpha\, ,
\label{e:rt}
\end{equation} 
where $\rho_0$ is the density at the time $\tau_0$. 
From this, the expression \eqref{rtau} can be recovered. 
This time-dependence of the density may not be realised during 
the whole evolution of the fireball; we assume it just in the final
stage of the collision. Therefore, the time $\tau$ is now {\em defined} 
by the dilution rate $\gamma$
\begin{equation}
\gamma \equiv 
-\frac{1}{\rho}\frac{\partial\rho}{\partial\tau} = \frac{\alpha}{\tau}\, .
\label{e:dil}
\end{equation}
Even though we are interested in the situation along the longitudinal
symmetry axis of the fireball, the time $\tau$ 
agrees with the ``usual'' longitudinal proper time $\sqrt{t^2 - z^2}$
only if the density evolves according to \eqref{e:rt} from the very beginning 
of the collision ($t=0$). Otherwise it corresponds to a different 
starting point $t_0$ of the time scale 
\footnote{Technically, if we were 
solving for the whole fireball evolution, we would have to match the 
prescription \eqref{e:rt} and its first time-derivative to the 
time-dependence of the density at earlier times. This is done by tuning
$\rho_0$ {\em and} $\tau_0$. Therefore, $\tau_0$ has to be a free
parameter, and is not necessarily equal to the longitudinal proper time 
measured from the moment of the first approach of the colliding nuclei.}
\begin{equation}
\label{taudef}
\tau = \sqrt{(t-t_0)^2 - z^2}\, .
\end{equation}
This is analogous to the Hubble time in cosmology which is defined as 
the inverse of the expansion velocity gradient. We insert this 
dynamical information in the usual parametrization 
of the expansion four-velocity
\begin{eqnarray}
\label{e:4vel}
     u^\mu &= & (\cosh\eta\, \cosh\eta_t(\tau,r)\, , \,
                  \cos\phi\, \sinh\eta_t(\tau,r)\, , \\
        & &       \sin\phi\, \sinh\eta_t(\tau,r)\, , \,
                  \sinh\eta\, \cosh\eta_t(\tau,r) ) \, .
\nonumber
\end{eqnarray}
In accord with relation \eqref{taudef}, the space-time rapidity is defined as
\begin{equation}
\eta = \frac{1}{2} \ln \left [ \frac{(t-t_0) + z}{(t-t_0) - z} \right ] \, ,
  \nonumber
\end{equation}
and $r,\, \phi$ are the standard radial coordinates used in the plane 
transverse to the beam. The transverse rapidity $\eta_t$
will be assumed to grow {\em linearly} with the 
radial coordinate.  If the particle number is conserved, which is 
a good assumption at the end of the hadronic phase, the dilution 
rate is related to the divergence of the velocity field 
\cite{Schnedermann:1992hp}
\begin{equation}
  \label{e:cont}
  -\frac{1}{\rho}\frac{\partial\rho}{\partial \tau} =
  -\frac{1}{\rho}\, u^\mu\partial_\mu\rho =
  \partial_\mu u^\mu\, .
\end{equation}
If we assume that $\eta_t(\tau,r) \propto r$, then eq.~\eqref{e:cont} 
links the four-velocity field \eqref{e:4vel} with the ansatz \eqref{e:rt}.
This specifies the time dependence of the transverse rapidity
\begin{subequations}
  \label{casea}
  \begin{eqnarray}
  \label{etata}
  \eta_t(\tau,r) & = & \chi \frac{r}{\tau}\, ,\\
  \label{chi}
  \chi & = & \frac{\alpha - 1}{2} \, .
  \end{eqnarray}
\end{subequations}
In the vicinity of $r=0$, the choice
$\chi = 1$ leads to quasi-inertial flow which corresponds 
to an asymptotic solution of the fireball hydrodynamics \cite{csoh}.
The values $\chi < 1$ and $\chi > 1$ stand for radially decelerating 
and accelerating flow profiles, respectively. This illustrates the 
phenomenological consequences of a specific choice of the exponent
$\alpha$. Relation \eqref{chi} allows to rewrite the opacity integral
\eqref{e:opac}
\begin{equation}
\label{e:opacity}
     \int_{\tau_0}^\infty \, d\tau \, \R(\tau) =
        \frac{\R_0 \tau_0}{\alpha -1} =
        \frac{\R_0}{2} \left ( \frac{\chi}{\tau_0} \right )^{-1}
        \, .
\end{equation}
According to \eqref{etata},
$\chi/\tau_0$ is the gradient of transverse rapidity at the 
time $\tau_0$. At $r=0$, it is to a good approximation equal to 
the transverse velocity gradient. This substantiates the statement
made above: a higher scattering rate can be compensated by 
a stronger transverse flow and still lead to the same escape probability. 
We note that it is
not the average flow velocity, but the local flow {\em gradient}
which determines the local density decrease and thus enters \eqref{e:opacity}.
In Appendix \ref{a1} we relate this result to the freeze-out criterion 
requiring the scattering rate to be smaller than the dilution rate
\cite{garpman}.

For a second simple dynamical model of the fireball evolution, we turn
to recent hydrodynamic simulations \cite{kolb}. These indicate that the 
transverse expansion at the freeze-out stage may be better described
by the ansatz
\begin{subequations}
\label{caseb}
\begin{eqnarray}
\label{etatb}
\eta_t(\tau,r) & = & \xi r \, ,\\
\label{xi}
\xi(\tau) & = & \mbox{const} \, ,
\end{eqnarray}
\end{subequations}
rather than by the expressions \eqref{e:rt} and \eqref{casea} used above. 
In this case, the variable $\tau$ is still given by the relation 
\eqref{taudef}, 
but corresponds now to the inverse longitudinal gradient of the expansion
velocity
\begin{equation}
  \tau = ( \partial_0 u^0 + \partial_3 u^3 )^{-1}\, .
  \nonumber
\end{equation}
The ansatz \eqref{caseb} implies a different time-dependence of the
density. From eq.~\eqref{e:cont}, one finds for the 
density at the center of the fireball 
\begin{equation}
\rho(\tau) = \rho_0 \frac{\tau_0}{\tau} \, 
\exp\left ( - 2 \xi ( \tau - \tau_0) \right )\, ,
\end{equation}
where $\rho_0$ is again the density at the time $\tau_0$. 
The corresponding opacity integral reads
\begin{equation}
\int_{\tau_0}^\infty \, d\tau \, \R(\tau) =
\R_{0} \, \tau_0 \, \exp(2\xi \tau_0)\,
\Gamma(0,2\xi\tau_0)\, ,
\label{e:op-const}
\end{equation}
where $\Gamma(a,x)$ is the incomplete 
$\Gamma$-function \cite{abram}
\begin{eqnarray}
\nonumber
\Gamma(a,x) & = & 
\int_x^\infty dt \, t^{a-1} \, e^{-t} \, ,\\
& = & 
\Gamma(a) - \int_0^x dt\, t^{a-1} \, e^{-t} \, .
\nonumber
\end{eqnarray}
Relation \eqref{e:op-const} is less transparent than \eqref{e:opacity}
but it shows the same qualitative feature: stronger scattering
can be compensated by transverse expansion and lead to the 
same escape probability.

\section{The scattering rate in a thermal model}
\label{sec4}

In the previous section, we studied the effect of expansion on
freeze-out. We derived expressions for the opacity integral which 
depend on the scattering rate at a fixed time $\tau_0$. 
Here we calculate this scattering rate and we study how it depends
on hadrochemistry and temperature. In particular, we determine the
scattering rate corresponding to the hadronic final states at SPS 
and RHIC.


The scattering rate for a test pion of momentum $p$   
due to interactions with particles 
of type $i$ and momentum $k$ can be written as \cite{Prakash,Ftacnik:1988ar}
 \begin{equation}
 \label{e:pirest}
     \frac{d\R_i(x,p)}{d^3k} =  \rho_i(x,k)\, \sigma_i(s)\,
     \frac{\sqrt{(s-s_a)(s-s_b)}}{2 \sqrt{m^2_\pi +
p^2}\sqrt{m_i^2+k^2}}\, .
 \end{equation}
Here, $\sigma_i(s)$ denotes the total cross-section
for collinear collision, 
\begin{eqnarray}
s_a & = & (m_i + m_\pi)^2\, ,\\
s_b & = & (m_i-m_\pi)^2\, ,
\end{eqnarray}
and the cms energy $s=s(k,p)$.

We assume that the distribution of scattering partners $\rho_i(x,k)$
is given by the equilibrium form. In this way we neglect the modification
of the distribution function due to decoupling of some particles. 
Such approximation was argued to cause only a small error on the escape 
probability in the practically relevant cases \cite{Sinyukov:2002if}.
Thus we write
 \begin{equation}
 \label{e:therm}
  \rho_i(x,k) = \frac{g_i}{(2\pi)^3} \, \left [
    \exp\left( (E_k-\mu_i)/T \right ) \pm 1 \right ]^{-1} \, ,
 \end{equation}
where $g_i$ is the degeneracy of the species, $T$ is the temperature,
$E_k = \sqrt{k^2 + m_i^2}$ and 
the chemical potential $\mu_i$ fixes the total density. 
This is known to provide a good description of the hadronic final
state at SPS and RHIC \cite{Tomasik:1999cq,Csorgo:2002ry}.

The total scattering rate is obtained by integrating the expression 
(\ref{e:pirest}) over momentum $k$ and summing over all species $i$
\begin{equation}
  \label{e:rtot}
     \R(x,p) =
       \sum_i \int\ d^3k\,\frac{d\R_i(x,p)}{d^3k}
       \, .
\end{equation}

All pion scattering rates are computed in the rest frame of the 
hadron gas.
We include pions, (anti)nucleons,  kaons, rhos, and (anti)deltas
as scattering partners. The total cross section 
of pion-baryon scatterings is parameterized as \cite{Bass:1998ca}
 \begin{eqnarray}
   \nonumber
     \sigma(\sqrt{s}) & = & \sum_{r}
        \frac{\la j_i,\, m_i,\, j_\pi,\, m_\pi\, || \, J_r,\, 
              M_r \ra(2S_r +1)}
             {(2S_i + 1)(2S_\pi +1)} \\
       &&  \times
       \frac{\pi}{p^2_{\rm cms}}
         \frac{\Gamma_{r\to \pi i}\Gamma_{\rm tot}}{%
           (M_r - \sqrt{s})^2 + \Gamma^2_{\rm tot}/4} \, .
 \label{e:sigma}
 \end{eqnarray}
This is the usual Breit--Wigner resonance formula where $M_r$ is the 
resonance mass and $\Gamma_{\rm tot}$ and $\Gamma_{r\to \pi i}$ are the 
total and the partial width for the given decay channel, respectively. 
Summation in (\ref{e:sigma}) runs over all relevant resonance states
listed in \cite{Bass:1998ca}. Momentum in 
the cms is denoted as $p_{\rm cms}$. The pre-factor takes care of proper 
counting of spin and isospin states. The square of  
Clebsch--Gordan coefficient
$\la j_i,\, m_i,\, j_\pi,\, m_\pi\, || \, J_r,\, M_r \ra$ assures
that the appropriate fraction of the resonance state is picked
in coupling of the isospin states of the two scattering partners.
Particle properties are taken from
\cite{Groom:in} except for higher excitations with large uncertainties
which are taken from \cite{Bass:1998ca}.
For scattering on both strange and non-strange 
mesons, the formula \eqref{e:sigma} is used again and
a 5~mb momentum-independent contribution is added to account for 
elastic processes.
The cross-section with baryons is taken to saturate at
30~mb for $\sqrt{s}$ above 3~GeV/$c$. This, however, has a negligible
effect on the calculation since high momentum particles are suppressed
in the thermal distribution.

To fix the hadrochemical composition of scattering partners at SPS and
RHIC, an estimate of the chemical potentials entering (\ref{e:therm})
is needed. We take the chemical potential for direct pions from data
on pion phase-space densities, and those for all other particle species
from ratios of $dN/dy$ at midrapidity. 
For SPS at $\sqrt{s}=17\, A\mbox{GeV}$ and RHIC at
$ \sqrt{s}=130\, A\mbox{GeV}$
we determine the chemical potentials for three different temperatures: 90,
100 and 120~MeV.

In more detail: the pion chemical potential at the SPS is obtained by
comparing the $m_\perp$-dependence of the average phase-space density
measured by NA44 \cite{Murray&Holzer} and WA98 \cite{wa98-psd} to
the result expected from a thermalized boost-invariant source with box
transverse density profile and a transverse flow profile
$\eta_t = \sqrt{2} \eta_f r/R_{\rm box}$ \cite{Tomasik:2001uz}.
The transverse momentum spectra relate transverse flow and temperature,
such that $\eta_f \simeq 0.7$ corresponds to $T =$
90 -- 100~MeV and $\eta_f\simeq0.55$ for $T=120$~MeV. From the ratios of
$dN/dy$ at mid-rapidity measured for pions, protons, antiprotons, positive
and negative kaons \cite{Bachler:1999hu,Afanasiev:2002mx}, particle
densities and corresponding chemical potentials are extracted under
the assumption that all particles originate from the same thermal
source, in accord with the use of equilibrated distribution in 
eq. \eqref{e:pirest}. Resonance decay contribution to pion
production is accounted for. The density of neutrons is assumed to
be the same as that of protons.
Chemical potentials for $\rho$'s, $\Delta$'s and $\bar \Delta$'s
are deduced by requiring detailed balance \cite{Bebie:1991ij}
\begin{equation}
     \mu_\rho =  2 \mu_\pi\, ,\quad
     \mu_\Delta =  \mu_p + \mu_\pi\, ,\quad
     \mu_{\bar \Delta} =  \mu_{\bar p} + \mu_\pi\, .
\end{equation}
The chemical potentials extracted in this way are summarized in
Table~\ref{t:SPSmus}. The upper and lower values of $\mu$'s were
chosen such that our parameters over- or under-predict the data by
at least the maximal amount allowed by the quoted error bars. This
ensures insensitivity of our conclusions against fine-tuning of
parameters.

%
\begin{table}[t]
\begin{tabular}{cr@{ -- }lr@{ -- }lr@{ -- }l}
temperature & \multicolumn{2}{c}{90 MeV} & \multicolumn{2}{c}{100 MeV} & 
\multicolumn{2}{c}{120 MeV}\\
\hline
$\mu_p$ & 477 & 497 & 435 & 455 & 350 & 379 \\
$\mu_{\bar p}$ & 301 & 321 & 238 & 259 & 114 & 143 \\
$\mu_\pi$ & 50 & 65 & 38 & 53 & 10 & 30 \\
$\mu_\Delta$ & 527 & 562 & 473 & 508 & 360 & 409 \\
$\mu_{\bar \Delta}$ & 351 & 386 & 276 & 312 & 124 & 173 \\
$\mu_\rho$ & 100 & 130 & 76 & 106 & 20 & 60 \\
$\mu_K$ & 162 & 182 & 132 & 153 & 72 & 102 \\
$\mu_{\bar K}$ & 111 & 131 & 76 & 96 & 4 & 33
\end{tabular}
\caption{Chemical potentials in units of MeV 
used in the calculation for SPS at
$\sqrt{s} = 17\, A\mbox{GeV}$.\label{t:SPSmus}}
\end{table}
%
\begin{table}[t]
\begin{tabular}{cr@{ -- }lr@{ -- }lr@{ -- }l}
temperature & \multicolumn{2}{c}{90 MeV} & \multicolumn{2}{c}{100 MeV} & 
\multicolumn{2}{c}{120 MeV}\\
\hline
$\mu_p$ & 442 & 489 & 393 & 452 & 291 & 373 \\
$\mu_{\bar p}$ & 407 & 458 & 354 & 418 & 245 & 332 \\
$\mu_\pi$ & 78 & 100 & 70 & 97 & 50 & 85 \\
$\mu_\Delta$ & 520 & 589 & 463 & 549 & 341 & 458 \\
$\mu_{\bar\Delta}$ & 485 & 558 & 424 & 515 & 295 & 417 \\
$\mu_\rho$ & 156 & 200 & 140 & 194 & 100 & 170 \\
$\mu_K$ & 194 & 242 & 170 & 229 & 120 & 202 \\
$\mu_{\bar K}$ & 179 & 230 & 153 & 217 & 100 & 187
\end{tabular}
\caption{Chemical potentials in units of MeV 
used in the calculation for RHIC at
$\sqrt{s}=130\, A\mbox{GeV}$.\label{t:RHICmus}}
\end{table}
The same procedure is repeated for RHIC at $\sqrt{s}=130~A\mbox{GeV}$.
Results
are summarized in Table~\ref{t:RHICmus}. The yields at mid-rapidity for
pions, (anti)protons, and kaons were taken from \cite{Adcox:2001mf}.
Unlike at SPS, the feed-down to (anti)proton yields from weak decays
was not corrected for in the used data. We performed a simple correction
by assuming that the yields of  
$\Lambda$'s and $\Sigma$'s are given by
the same temperature and their chemical potentials equal to that of the 
protons. The preliminary
data on pion phase-space density presented recently by STAR
show small statistical error bars only \cite{lray}, and lie close to
the upper bound of the large systematic errors quoted earlier
\cite{cramer}. Our analysis accounts for this not fully
clarified experimental situation by associating to the central
value of the most recent data \cite{lray}
the large (asymmetric) systematic errors of \cite{cramer}.
To avoid a bias in relating SPS and RHIC energies, we
compare the upper bound of the chemical potential for
RHIC in Table \ref{t:RHICmus} (which corresponds to the central
value in \cite{lray}) to the upper bound
listed in Table~\ref{t:SPSmus} for SPS. We thus underestimate
the difference between RHIC and SPS.

%
\begin{figure}[t]
\vbox{
   \begin{center}
      \epsfig{file=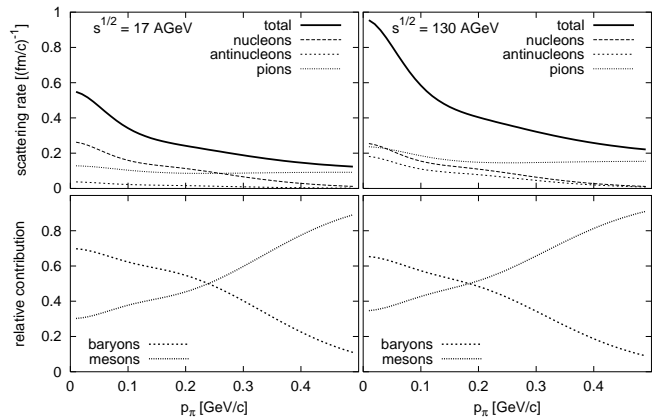,scale=0.686}
   \end {center}
\caption{Pion scattering rate as a function of pion momentum
at $T=100\, \mbox{MeV}$ and
the highest possible values of  chemical potentials for SPS (left column)
and RHIC (right column). Contributions to the total scattering rate
from scattering on nucleons, antinucleons and pions are indicated.
The lower row shows the baryonic and mesonic relative contributions.
\label{F:contribs}}
}
\end{figure}
%
%
\begin{figure}[t]
\vspace*{0.4cm}
\vbox{
   \begin{center}
      \epsfig{file=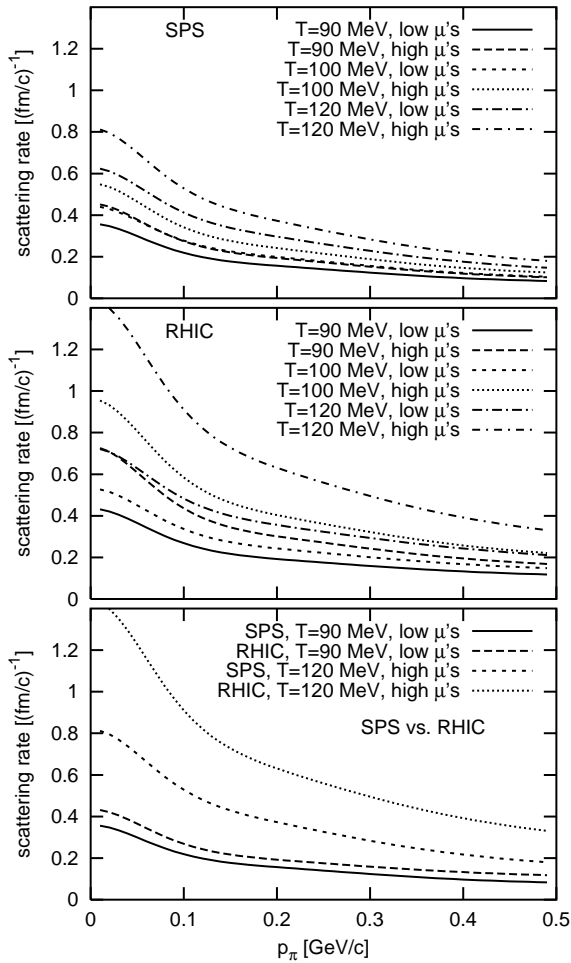,scale=1.5}
   \end {center}
\caption{Pion scattering rate as a function of pion momentum
calculated for temperatures 90, 100, and 120 MeV and all corresponding
sets of chemical potentials from Tables~\ref{t:SPSmus} (SPS at
$\sqrt{s}=17~A\mbox{GeV}$) and \ref{t:RHICmus} (RHIC at
$\sqrt{s}=130~A\mbox{GeV}$). Lowest panel summarizes
the lowest and the highest curves from both SPS and RHIC.
\label{F:comp}}
}
\end{figure}

Figure~\ref{F:contribs} shows the corresponding scattering rates
for $T= $ 100~MeV. The smaller contribution of nucleons
at RHIC is largely compensated by
antinucleons. Despite the increased pion phase-space
density at RHIC, the pionic contribution does not dominate
the total scattering rate because of the small pion-pion
cross section. Scattering on pions is of comparable importance to
scattering on nucleons and antinucleons: both particle species
lead to a momentum averaged scattering rate of
$\sim 0.2\,(\mbox{fm}/c)^{-1}$.

We observe that the increase of pion phase-space density at 
RHIC when compared to SPS \cite{lray,cramer} has no significant 
impact on the scattering rate.
The earlier suggestion that the pion phase-space 
density at the freeze-out should be a universal quantity 
\cite{Ferenc:1999ku} did not account for the contribution
of other particle species to the pion scattering rate. 

Figure~\ref{F:comp} summarizes our results for the various sets of
freeze-out temperature and chemical potentials which limit the
range of values consistent with data from SPS and
RHIC. Although particle densities are approximately the same for
all temperatures, the scattering rate increases significantly
with temperature. On the other hand we see a generic decrease of 
the scattering rate with the pion momentum. This behavior 
is mainly dictated by phase-space available for the collisions, but
also depends on the cross-section as a function of $s$. We
discuss this behavior in more detail in Appendix \ref{a2}.

In the lowest panel of Fig~\ref{F:comp} we compare the extreme results 
for SPS and RHIC. If the scattering rate in one of the systems
was clearly larger than in the other one, this would indicate 
a stronger transverse expansion in that system, as follows from  
Section~\ref{sec3}. We see that the present data do not allow 
to make any conclusion on this subject just by using the scattering rate.

\section{Conclusions}
\label{sec5}

In this paper, we have 
have calculated the scattering rates of pions which are 
characteristic for the late hadronic phase of the collisions at 
SPS and RHIC. We also illustrated for simple 
examples how these scattering rates determine the escape probability,
and how the escape probability is affected by velocity gradients. 
Moreover, we found that the temperature dependence of the scattering 
rate allows to constrain the range of temperatures at which
pion emission is significant. To estimate this temperature range, 
we combined the scattering rates of Section \ref{sec4} with the 
expressions for the opacity integral \eqref{e:opacity} and 
\eqref{e:op-const}. Here, we discuss what is needed to overcome
the limitation of our approach and what our results imply for 
freeze-out in the realistic case.

To go beyond the calculations presented here will require
a refined dynamical model which follows the full space-time
evolution of the fireball. Most likely, this will involve
numerical simulations. In particular, one may hope to extend 
in this way the expressions for the opacity integral in Section 
\ref{sec3} to the case of particles with finite momentum. 
For a realistic scenario, we envisage the following
additional effects: 
\begin{enumerate}
\item
\label{i1}
In a transversely expanding fireball, particles with $\pt > 0$
are produced from parts of the system that co-move transversely.
Their momentum relative to the medium $p_\pi$ is thus smaller than
$\pt$. The corresponding scattering rate $\R(p_\pi)$ is thus higher 
than the naively expected value at $\pt$. Hence, this effect
{\em lowers} the escape probability.
\item
\label{i2}
A particle with finite $\pt$ can escape the fireball region in
a finite time. Thus, the corresponding opacity integral ~\eqref{e:escprob}
receives a contribution from a finite time only, in contrast to the
infinite time which is accumulated by a test particle with $\pt = 0$
in our simplified calculation. This effect leads to a lower
value of the opacity integral and {\em enhances} the escape 
probability.
\end{enumerate}

Also, the evaluation of the general freeze-out criterion \eqref{e:ufc} 
is complicated and requires a complete knowledge of the space-time 
evolution of the fireball. To gain some understanding of the freeze-out 
process from the models studied here, however, it is instructive
to turn to a simpler condition. Let us  
define freeze-out for particles at position $x$ with momentum $p$ 
to occur if the corresponding escape probability $\P(x,p)$ increases 
above a certain threshold value. Doing this, we actually specify a 
three-dimensional hypersurface on which a given fraction 
$f_{\rm esc}/f = \P$ of all particles is already decoupled.
However, in contrast to previous freeze-out criteria 
\cite{ceres,Ferenc:1999ku,Alber:si}, this condition does not only 
depend on the medium, but also on the particle momentum. Moreover, 
we can access some properties of the whole four-dimensional 
freeze-out region by varying the threshold value for $\P$.
In this way, the above model calculations give some insight into 
freeze-out criteria which are more general than the Cooper-Frye type.
To put in numbers: A realistic estimate of the transverse flow gradient 
$\xi = 0.08 \, \mbox{fm}^{-1}$ and the Bjorken freeze-out time 
$\tau_0 = 10\, \mbox{fm}/c$ is taken from \cite{Tomasik:1999cq}.
The value $\xi = 0.08 \, \mbox{fm}^{-1}$ is also in agreement 
with hydrodynamic simulations \cite{kolb}. In order to achieve 
the escape probability of at least 1/3, the opacity integral must be smaller 
than $-\ln 1/3 \approx 1.1$. Then, eq.~\eqref{e:opacity}
leads to a scattering rate of less than $0.18\, (\mbox{fm}/c)^{-1}$ while 
eq.~\eqref{e:op-const} puts the upper limit for the scattering
rate to $0.26\, (\mbox{fm}/c)^{-1}$. At a temperature of 120~MeV, such a
low value of the scattering rate is possibly reached at RHIC for
$p_\pi > 0.4\, \mbox{GeV}/c$ and at SPS at $p_\pi > 0.3\, \mbox{GeV}/c$,
as seen in Fig.~\ref{F:comp}. If we increase the required escape probability
to 1/2, the scattering rate must be smaller 0.11 or 0.16,
according to eqs. \eqref{e:opacity} and \eqref{e:op-const}, respectively.
For particles with $p_\pi<0.4 \, \mbox{GeV}/c$, the temperature has to drop 
to 100~MeV if $\P$  should reach 1/2. Particles
with $p_\pi < 0.2\, \mbox{GeV}/c$ must wait even longer. On the other hand,
if we put the required value of $\P$ to 0.1, eq.~\eqref{e:op-const}
leads to an upper bound of $0.54  \, (\mbox{fm}/c)^{-1}$ for the 
scattering rate. In the case of $T=120\, \mbox{MeV}$ and high
$\mu$'s at RHIC pions with $p_\pi > 0.25\, \mbox{GeV}/c$ fulfill
this condition; in the less extreme cases, the 10\% escape probability
is reached by almost all pions.

This illustrates that particles decouple from the fireball gradually.
It indicates that $\approx$ 10\% of the pions are decoupled at 
$T=120\, \mbox{MeV}$, but approximately half of them will escape
at local temperatures below 100~MeV. Thus, a large fraction of the 
particles decouples at rather low temperatures. We note that
a temperature of 120~MeV was assumed as a freeze-out temperature
in a recent work \cite{ceres} where a pion mean free path 
$\approx 1\, \mbox{fm}$ was suggested as the universal freeze-out criterion.
This work neglects the temperature dependence of the escape 
probability and its dependence on the momentum of the particle.
Also, it does not consider the possible cancellation between 
stronger scattering and stronger transverse expansion. 
In our approach, a thermal distribution of $T=120\, \mbox{MeV}$ at RHIC 
results in a pion mean free path of only 1.7~fm, only slightly larger
than \cite{ceres}. However, only 10 \% of all particles are decoupled
at this temperature. At lower 
temperatures, where the fraction of decoupled particles 
reaches 50\%, we find a significantly larger average pion mean 
free path of at least 3--5~fm. 

Let us further comment on the significance of the momentum dependence
of the scattering rate. As discussed earlier, at certain $\pt$ the 
particle has a momentum with respect to the surrounding medium $p_\pi < \pt$,
and thus the $\pt$-dependence of the scattering rate can be flatter than 
what is plotted in Fig.~\ref{F:comp}. We expect, however,
that the monotonic
increase of the escape probability with $\pt$ will be robust
against refinements of our calculation. This suggests that
in contrast to the idealization of a sharp freeze-out along a 
three-dimensional hypersurface, particle escape is ordered in
momentum with high-$\pt$ particles freezing-out earlier and
thus originating from a smaller fireball. This effect 
could lead to a stronger transverse mass dependence of HBT radii 
\cite{Grassi:2000ke} than what is found in current model studies
based on a Cooper-Frye freeze-out condition. These latter models  
(see e.g. the {\em blast-wave model} discussed in \cite{moriond})
have difficulties in reproducing the observed strong 
$M_\perp$-dependence of HBT radii at RHIC. In what concerns the 
mechanism proposed here, one may envisage compensating
effects: for example, if the transverse flow gradient is stronger at
later times, then this will reduce the homogeneity regions measured 
by HBT radii at later times. Such a scenario seems unlikely since
it implies acceleration of the collective expansion at a time when 
particles slowly cease to interact. However, to substantiate this
expectation, a model which fully implements the fireball dynamics
is needed. 
In our opinion, this warrants further investigations.  In particular,  
hydrodynamical simulations should be revisited, 
which are typically based on sharp 3-dimensional
freeze-out and thus miss any contribution from freeze-out along
finite four-volumes.

We thank J.G.~Cramer, U.~Heinz, J.~Pi\v s\'ut, and K.~Redlich for fruitful
discussions.

\appendix

\section{Interplay of the scattering rate and the dilution rate}
\label{a1}

Here, we give further details on the relation between the 
dilution rate and the scattering rate. 
We start from the opacity integral \eqref{e:opac} and we express the 
exponent $\alpha$ in terms of the dilution rate 
$\gamma$ via eq.~\eqref{e:dil}. This leads to
\begin{equation}
\int_{\tau_0}^\infty d\tau \, \R(\tau) = \frac{\R_0 \tau_0}{\alpha - 1} 
= \frac{\R_0}{\gamma - \frac{1}{\tau_0}} \, .
\label{e:rates}
\end{equation}
This can be directly compared to the statement \cite{garpman} that 
freeze-out happens when the dilution rate is at least as big as the 
scattering rate. For the scenario of continuous freeze-out considered
here, a reasonable escape probability is reached when the opacity 
integral is of order one or smaller.
In the model of eqs.~\eqref{rtau} and \eqref{e:rate} this
leads to the condition 
\[
\gamma - \frac{1}{\tau_0} \gtrsim \R_0 \, ,
\]
i.e., there is a factor $-1/\tau_0$ in addition to the standard 
criterion $\gamma \gtrsim \R_0$ \cite{garpman}. The 
details of the relation between the scattering rate and the dilution rate
at the freeze-out will depend on the particular time evolution 
of $\R$ and $\rho$, but the feature that larger $\R$ can be
compensated by larger $\gamma$ is generic.


\section{General expression for the scattering rate}
\label{a2}

In this appendix we derive a general expression for 
the scattering rate. This allows us to investigate its dependences 
on temperature and the test particle momentum.

Following eq.~\eqref{e:rtot} we write the scattering for some particle
species of mass $m$. Assuming a Boltzmann 
distribution of the scattering partners and suppressing the position 
dependence, 
\begin{multline}
\label{ea:rtot}
{\cal R}(p) 
 =  \int \frac{d^3k}{(2\pi)^3} \, g
\, \exp\left(-\frac{E_k - \mu}{T}\right ) \\
\times \sigma(s)\, 
\frac{\sqrt{(s-s_a)(s-s_b)}}{2 \sqrt{m^2_\pi + p^2}\sqrt{m^2+k^2}}
\, .
\end{multline}
If we change the integration 
variables to the center of mass energy $s$ and the
energy of the scattering partner $E_k$, and perform the $E_k$-integration,
this integral leads to
\begin{multline}
\label{ea:R-long}
 {\cal R}(p)  =  \frac{g \, e^{\mu/T}}{8\pi^2\, p\sqrt{p^2 + m_\pi^2}} \\
  \times 
  \int_{(m_\pi + m)^2}^\infty ds\, \sigma(s) 
  \sqrt{(s-m_\pi^2-m^2)^2 - 4m_\pi^2 m^2}\\
  \times \sinh\left( \frac{p}{2 T m_\pi^2} 
  \sqrt{(s-m_\pi^2-m^2)^2 - 4m_\pi^2 m^2} \right )\\
  \times
  \exp\left ( - \frac{\sqrt{p^2+m_\pi^2}}{2 T m_\pi^2} 
               (s - m_\pi^2 - m^2) \right )\, .
\end{multline}
This scattering rate can be expressed as a convolution of the
cross-section with the {\em ``distribution of two-particle approaches''} 
${\cal D}(p,s)$
\begin{equation}
\label{ea:ddef}
{\cal R}(p) = \int_{(m_\pi + m)^2}^\infty {\cal D}(p,s)\, \sigma(s)\, ds \, .
\end{equation}
By comparing with \eqref{ea:R-long}, we obtain
\begin{multline}
\label{ea:dexpr}
  {\cal D}(p,s) = \frac{g \, e^{\mu/T}
                   \sqrt{(s-m_\pi^2-m^2)^2 - 4m_\pi^2 m^2}}
                   {8\pi^2\, p\sqrt{p^2 + m_\pi^2}} \\
  \times 
  \sinh\left( \frac{p}{2 T m_\pi^2} 
  \sqrt{(s-m_\pi^2-m^2)^2 - 4m_\pi^2 m^2} \right )\\
  \times
  \exp\left ( - \frac{\sqrt{p^2+m_\pi^2}}{2 T m_\pi^2} 
  (s - m_\pi^2 - m^2) \right ) \, .
\end{multline}
In the arguments of the exponential functions of \eqref{ea:dexpr},
the momentum of the test particle and the temperature are inversely
related. Thus if one ignores the $1/p\sqrt{p^2+m_\pi^2}$ pre-factor, 
increasing $T$ has the same effect as decreasing $p$. This offers an 
explanation why $\cal R$ increases with raising the temperature but decreases 
when the momentum becomes larger. There 
is no collision dynamics going into the calculation of $\cal D$; it
represents merely the phase-space populated with thermally distributed 
scattering partners.

However, the temperature and momentum dependence of the scattering
rate is not given solely by the distribution of two-particle approaches,
but also reflects details of the convolution of ${\cal D}(p,s)$ with 
the scattering cross section $\sigma(s)$. This can be seen e.g.\ by
comparing our results with those obtained in \cite{Prakash}. In that 
paper, the authors used a simple prescription 
$\sigma(s) \propto \delta(s - M^2_r)$ which leads to a qualitatively 
different dependence of the scattering rate on the test particle momentum. 
We verified that this is a consequence of neglecting the 
width of the resonance. By only assuming the lowest resonance states
in a given channel and narrowing the resonance shape, we were able to 
reproduce the results of \cite{Prakash} as a limiting case of our 
calculation. 


\end{document}